\documentclass[11pt,a4paper]{article}
\usepackage{a4wide}
\RequirePackage{amsthm}

\usepackage{amsfonts}
\usepackage{pgfplots}
\usepackage{hyperref}
\usepackage{environ}
\pgfplotsset{compat=newest}
\pgfplotsset{plot coordinates/math parser=false}
\usepackage{mathtools}
\usepackage{mathrsfs}
\usepackage{float}
\usepackage{caption}
\usepackage{subcaption}
\usepackage{algorithm}
\usepackage{algpseudocode}

\title{Source Identification by Consensus-Based Optimization}
 
\author{Jan Friedrich\footnote{Chair of Numerical Analysis, Institute for Geometry and Applied Mathematics, RWTH Aachen University, Templergraben 55, 52056 Aachen, Germany}, Sarah Schraven\footnote{Institute for Experimental Molecular Imaging, RWTH Aachen University, Forckenbeckstr. 55, 52074 Aachen, Germany}, Fabian Kie{\ss}ling\footnotemark[2], Michael Herty\footnotemark[1]}

\begin{document}
\maketitle

\begin{abstract} 
A consensus-based optimization (CBO) algorithm, which enables derivative and mesh-free optimization, is presented to localize a bioluminescent source.
The light propagation is modeled by the radiative transfer equation  approximated by spherical harmonics.
The approach is investigated for a hierarchy of simplified diffusion models in simulated environments and tissue-mimicking phantoms.
In simulations, the state-of-the-art diffusive approximation gives reliable results for heavily scattering media. However, higher-order models achieve better localization and more accurate source intensities for deeper sources and in the presence of artificial noise in strongly absorbing, but only moderately scattering media.
In phantoms, higher-order models give lower approximation errors and the most accurate localization, even for a high scattering coefficient.
These results demonstrate the potential of CBO to render higher-order models at lower computational cost while ensuring accurate localization in bioluminescence tomography.
\end{abstract}

\section{Introduction}
Bioluminescence imaging (BLI) is a widely used method for tracking cells, either to monitor tumor or pathogen growth and for assessing gene expression with reporter genes \cite{sadikot2005bioluminescence}. Usually, image acquisition is fast and has low background noise due to its high specificity.  However, to find the source of light, 3D tomographic reconstruction models need to be applied, which require coupling with anatomical and optical information, as well as numerical simulation to compute the diffuse light propagation from the source of light to the camera.\\ 
The radiative transfer equation (RTE) \cite{case1968linear} is a well-known model for the propagation of light in heterogeneous media, and,  optical tomography problems have been considered using the RTE, e.g., in the mathematical community \cite{arridge2009optical}. Its numerical approximation usually follows a semi-discretization applied in the angular direction with two major approaches:  discrete ordinate methods and moment methods.
Discrete ordinate methods apply quadrature formulas on the sphere  \cite{case1968linear,johnson1983convergence} and those methods are closely related to Galerkin methods \cite{wareing2001discontinuous,kophazi2015space}.
Iterative solvers are used to solve the resulting set of algebraic equations \cite{adams2002fast}.
On the other hand, moment methods are typically based on truncated spherical harmonic expansions using variational discretization strategies \cite{pitkaranta1977approximate,egger2012mixed} or, for the time-dependent RTE, finite difference methods \cite{seibold2014starmap}.\\
In applications, a diffusive approximation (DA) of the RTE is often employed if the scattering is dominant.
The latter approach is essentially a first-order moment method, also referred to as $SP_1$, and has been used in locating light sources in BLI \cite{gu2004three,chaudhari2005hyperspectral,jiang2007image,han2008bioluminescence,kuo2007three,liu2022multispectral}.
In addition, higher-order moment methods have been considered in \cite{lu2009spectrally,yang2015performance,liu2010evaluation,lu2009experimental,guo2013improved,kokhanovsky2013multi,klose2010vivo} and recently in combination with neural networks \cite{yu2021bioluminescence}.
These methods have also been used in other inverse problems related to RTE, such as in radiotherapy \cite{frank2008optimal,herty2007asymptotic}.
In the majority of cases, higher-order moment methods demonstrate superior performance and broader applicability compared to the DA for various types of tissues \cite{yang2015performance}.
This results in a more precise reconstruction of the light source.
Due to the trade-off between accuracy and computational cost, a third-order moment approximation has proven to be a highly effective approach in bioluminescence tomography (BLT) involving both homogeneous and heterogeneous phantoms, e.g. in \cite{liu2010evaluation,lu2009spectrally}.
These approximations yield transport-like solutions at lower computational costs compared with higher-order moment methods.
We refer to \cite{qin2014recent} for an overview on BLT.
A standard approach to solve the inverse problem is to use an adjoint calculus to obtain gradients. This calculus requires an approximation of the adjoint RTE or approximations thereof. We refer to the literature for further details \cite{arridge2009optical, gu2004three, jiang2007image, lu2009spectrally, lu2009experimental, frank2008optimal, herty2007asymptotic}. Here, we propose a different approach not relying on gradients. This allows non-differentiable cost functions as well as different higher-order moment models to be considered during the optimization.\\
Hence, the objective of this study is to propose an alternative optimization algorithm for BLT: a so-called consensus-based optimization (CBO) algorithm.
These algorithms are derivative-free and solve the optimization problem directly, such that they are mesh-free, i.e. the number of iteration steps does not depend on the underlying discretization.
More importantly, they require only a forward model to solve the RTE.
Hence, the algorithm can be easily employed with different moment hierarchies and it allows to (numerically) increase the number of moments in an adaptive manner, i.e. starting with the DA and increasing the number of moments when the minimization point is approached. 
This keeps computational costs lower and achieves higher accuracy (higher moment approximation) than using a fixed number of moments during the optimization process.

\section{Mathematical modeling}

\subsection{Radiative transfer equation}
 Light propagation in heterogeneous media is described by RTE with an (unknown) isotropic source term $q(x)$, i.e., the light source.
\begin{equation}\label{eq:RTE}
  \begin{aligned}
          \Omega \nabla_x \psi (x,\Omega)+\Sigma_t(x) \psi(x,\Omega)
=&\int_{S^2} s(x,\Omega\cdot \Omega')\psi(x,\Omega')d\Omega'+\frac{q(x)}{4\pi}
  \end{aligned}
\end{equation}
The angular flux $\psi \in \mathbb{R}$ describes the (light) particle density at space coordinate $x\in \mathcal{P} \subset \mathbb{R}^3,$ $\mathcal{P}$ being a convex phantom, and traveling in direction $\Omega\in S^2$, where $S^2$ is the unit sphere in 3D.
The particles undergo scattering and absorption in the heterogeneous medium. The parameters are defined as follows: $\Sigma_t:=\Sigma_{s}+\Sigma_a$ is the total cross section with the absorption cross section $\Sigma_a$ and scattering cross section $\Sigma_s$. 
The scattering kernel $s$ is normalized and for an isotropic scattering given by 
$s(x,\nu)=\frac{1}{4\pi}\Sigma_s(x)$.
The scattering and absorption coefficients are given by $\Sigma_a(x)=\mu_a(\lambda)$ and $\Sigma_s(x)=\mu_s'(\lambda)$,
where $\lambda$ is the wavelength and the scattering and absorption coefficients depend on the material of the phantom.
Boundary conditions for $\psi$ are given by vacuum boundary conditions due to the setup described below, i.e.,  $\psi(x,\Omega)=0$ for $(x,\Omega) \in \partial \mathcal{P}\times S^2$ and $n(x)\cdot \Omega<0$, where $n$ is the normal vector.

\subsection{Hierarchical Simplified Moment Approximation}
The RTE is posed in a {6D} phase space (space $x$ and direction $\Omega$) rendering the simulation and subsequent optimization computationally expensive. Therefore, reduced models using a series expansion in the angular dependence with spherical harmonic functions are proposed~\cite{brunner2005two,seibold2014starmap,MR2769988}. 
The resulting spherical harmonic equations, called $P_N$ equations, can be further simplified in the case of scattering media \cite{Gelbard1,Gelbard2,pomraning1993asymptotic,larsen1993asymptotic}. 
Here, we follow the ad hoc approach considered in \cite{Gelbard1,Gelbard2,olbrant2013asymptotic}. 
For the computational results $SP_1$, $SP_3$ and $SP_5$ approximations will be considered. The $SP_5$ equations are given by:
\begin{align}\nonumber
    &-\nabla_x \left(\frac{1}{\Sigma_t(x)}\nabla_x \left(\frac{1}{3}\phi_0+\frac{2}{3}\phi_2\right)\right)+\Sigma_a(x)\phi_0(x)=q(x)\\ \label{eq:SPN}
    &-\nabla_x \left(\frac{1}{\Sigma_t(x)}\nabla_x \left(\frac{2}{15}\phi_0+\frac{11}{21}\phi_2+\frac{12}{35}\phi_4\right)\right)+\Sigma_t(x)\phi_2(x)=0\\
    &-\nabla_x \left(\frac{1}{\Sigma_t(x)}\nabla_x \left(\frac{12}{143}\phi_2+\frac{39}{133}\phi_4\right)\right)+\Sigma_t(x)\phi_4(x)=0 \nonumber
\end{align}
The $SP_1$ equation, usually called DA, is obtained by only considering the first equation and setting $\phi_2=\phi_4=0$.
Analogously, the $SP_3$ equations are given by the first two equations and $\phi_4=0$.
The zeroth moment $\phi_0$ of each $SP_N$ approximation is also called the scalar flux.
Vacuum boundary conditions are enforced by a perfectly matched layer approach \cite{egger2019perfectly}, i.e.,  the domain around the convex phantom $\mathcal{P}$ is extended by an absorbing layer. By selecting a sufficiently large artificial absorption in the surrounding layer, the consistency error introduced by this approach can be made small enough to obtain an appropriate solution of the $SP_N$ equations.

\subsection{Inverse Problem}
Using the $SP_N$ equations, we identify a bioluminescent light source $q$ in a phantom by  given reference measurements (data) $U_{ref}(\lambda)$ on a part of the boundary of the probe $\mathcal{S}\subset \mathcal{\partial P}$. 
Depending on the equipment available, the surface data can be recorded from one or more perspectives, such that $U_{ref}$ can be available on the whole domain $\partial \mathcal{P}$ or on a subset.
Further, bioluminescence images are available for a selected number of $k$ wavelengths $\lambda$.\\
Let us denote by $q_i,\ i=1,\dots,k$ the sources for wavelength $i$ and by $Q_{ad}$ the admissible set of sources.
We assume $Q_{ad}$ to be a closed and convex subset of $L^{2}(\mathcal{P})$. Denote by 
$\mathbf{q}=(q_1,\dots,q_k)^T$ and by $\mathbf{Q}=\{\mathbf{q}: q_i \in L^2(\mathcal{P}),\ i=1,\dots,k\}$ and analogously $\mathbf{Q}_{ad}$. For more details, we refer to \cite{han2007theoretical}.\\
Let the scalar flux for a wavelength $\lambda$ be given by $\phi_0(\cdot)=\int_{S^2}\psi(\cdot,\Omega')d\Omega'$ denoted in the following by $\phi_0(\cdot,\lambda,q)$.
Here, $\phi_0$ is obtained as a solution to the moment approximation given in \eqref{eq:SPN}.
Then, we define a Tikhonov regularization functional \cite{engl1996regularization} by
\begin{align}
f_\eta(\mathbf{q}):=\sum_{i=1}^k\frac{\| U_{ref}(\lambda_i) - \phi_0(\cdot,\lambda_i,q_i)\|^2_{L^2(\mathcal{S})}}{\| U_{ref}(\lambda_i)\|^2_{L^2( \mathcal{S})}}+\eta \| q_i\|^2_{L^2(\mathcal{P})}.
 \end{align}
For identifying the light source ${\bf q}$ we pose the following inverse problem: 
\begin{align}\label{eq:inverseproblem}
   \text{Find }\mathbf{q}_\eta\in \mathbf{Q}_{ad} \text{ such that } f_\eta(\mathbf{q}_\eta)=\min_{\mathbf{q}\in \mathbf{Q}_{ad}}f_\eta(\mathbf{q}).
\end{align}
The previous formulation holds for $x\in \mathbb{R}^3$ and data $U_{ref}$ available at the surface of the probe.
Similar mathematical problems have been studied in \cite{gu2004three,chaudhari2005hyperspectral,jiang2007image,han2008bioluminescence,kuo2007three} for the DA ($SP_1$) and \cite{lu2009spectrally} for a third-order moment approximation.
Without a regularization term (i.e., $\eta=0$), problem \eqref{eq:inverseproblem} is generally ill-posed \cite{han2006mathematical,han2007theoretical,han2008bioluminescence}.
However, when $\eta>0$, a unique solution exists \cite{han2006mathematical,han2007theoretical}.
Further, if $Q_{ad}$ is bounded and $\eta=0$, the existence of at least one solution of \eqref{eq:inverseproblem} is guaranteed \cite{han2006mathematical,han2007theoretical}.\\
In this work, two types of sources, independent of the wavelengths, are considered: a sphere and a quadrilateral --  a not self-intersecting four-sided polygon. We assume that each source lies completely in (a convex subset) of $\mathcal{P}$. Without loss of generality, the quadrilateral is assumed to be parallel to the $x_1-x_2$ plane.
Note that the functions $q_i,\ i=1,\dots,k$ are described by $9+k$ real parameters, the corners of the quadrilateral at a certain depth, both fixed for all wavelengths and the intensity values. Further, they are described by $4+k$ real parameters in case of a sphere. Additionally, the unknown intensity values are assumed to be positive, but bounded. The set of the parameters is denoted by $\mathbf{X}_{ad}$.
The set of source functions $q_i$ generated by $X_{ad}$ is not necessarily convex for the cases considered.
Nevertheless, \eqref{eq:inverseproblem} can be equivalently stated in parameter space as
\begin{align}\label{eq:Opti2}
 &\min_{X\in\mathcal{X}_{ad}} f_\eta(X):=\sum_{i=1}^k\frac{\| U_{ref}(\lambda_i) - \phi_0(\cdot,\lambda_i,q_i(X))\|_{L^2(\mathcal{S})}}{\| U_{ref}(\lambda_i)\|_{L^2(\mathcal{S})}}+\eta \| q_i(X)\|_{L^2(\mathcal{P})}\\
 &\text{ subject to \eqref{eq:RTE} or \eqref{eq:SPN} for each }i=1,\dots,k.\nonumber
\end{align}
Since the source is continuous with respect to $X$, the existence of global minima is guaranteed.

\section{Optimization Algorithm}
\subsection{Consensus-based Optimization}
The inverse problem \eqref{eq:inverseproblem} is solved by finding the optimal value of equation \eqref{eq:Opti2}. Since this is a non-convex optimization problem,  a CBO algorithm \cite{pinnau2017consensus,borghi2023constrained,bae2022constrained,fornasier2022convergence} is employed.
This algorithm belongs to the class of stochastic particle optimization methods. 
Stochastic particle optimization methods have been used in BLT together with $SP_3$ approximations in \cite{kokhanovsky2013multi}.
We consider $i=1,\dots,M$ particles $X^i \in \mathcal{X}_{ad}\subset \mathbb{R}^d$.
The system of  particles evolves in pseudo-time $\tau$ following a stochastic differential equation, see e.g. \cite{pinnau2017consensus}:
\begin{align*}
    dX^i(\tau)=-\delta(X^i(\tau)-c_f^{\alpha}(X))d\tau+\sigma \text{diag} (X^i(\tau)-c_f^{\alpha}(X))dW^i(\tau).
\end{align*}
In case particles leave the domain $\mathcal{X}_{ad}$, they are projected back by $X^i(t)=\text{argmin} \{\|X^i(t)-y\|, y\in\mathcal{X}_{ad}\}$ as in \cite{bae2022constrained}.
The dynamics above are driven by two components: A drift part with strength $\delta\geq 0$ moves the particles towards the so-called consensus point $c_f^\alpha(X)$, which is given by
\begin{align*}
    c_f^\alpha(X)=\frac{\sum_{i=1}^MX^i(t)\omega_\alpha^f(X^i(t))}{\sum_{i=1}^M\omega_\alpha^f(X^i(t))},\quad \omega_\alpha^f(x)=\exp(-\alpha f(x)),
\end{align*}
and the anisotropic diffusion part allows for the exploration of $\mathcal{X}_{ad}$ by using the Brownian motion $W^i$ and $\sigma\geq0$. 
These two components allow the particles to form a consensus around $c_f^\alpha(X)$ such that $\frac{1}{M}\sum_{i=1}^M\| X^i-c_f^\alpha(X)\|_2\to 0$ for $\tau \to \infty$.
Under mild assumptions on $f(\cdot)$, it can be shown  that $\frac1M \sum_{i=1}^M X^i(\tau) \to X^*$ for $\tau \to \infty$ and $\alpha$ sufficiently large, i.e. the expected value of the particles converges towards the global minimizer $X^*$ of \eqref{eq:Opti2}, see \cite{fornasier2022convergence,bae2022constrained}.\\
The CBO method is gradient-free, thereby allowing the consideration of alternative cost functions in a straightforward manner,  e.g., the $L^1$-norm as in \cite{guo2013improved,chu2023graph} or  the $L^{1/2}$-norm as in \cite{yu2018source}.
As a first-optimize-and-then-discretize approach, CBO directly solves the underlying inverse problem and avoids the computation of adjoints.
The method is also mesh-free, whereby the number of iteration steps required for a consensus to be formed is independent of the computational mesh employed. 
To evaluate the objective function \eqref{eq:Opti2}, it is necessary to compute a solution to the RTE \eqref{eq:RTE}. 
Our approach is to utilize the hierarchy of $SP_N$ approximations, where 
the solution of the $SP_N$  can be computed in parallel for each particle.\\
Instead of using a fixed $N$ in the moment approximation during the optimization algorithm, as in \cite{lu2009spectrally,yang2015performance,liu2010evaluation,lu2009experimental,guo2013improved,kokhanovsky2013multi},  we use the underlying moment hierarchy (at least numerically) in the following manner:
As long as the particles do not concentrate below a certain threshold, a low-order moment approximation is used to minimize the computational cost, e.g. $SP_1$.
As the particles approach each other and hence the consensus point, the moment approximation is increased, e.g. to $SP_3$.
The algorithm terminates when the particles concentrate at the consensus point.  This approach results in a reduction of computational cost compared to the use of a fixed number of moments while maintaining or even exceeding the accuracy of the former.
A pseudo-code of the algorithm is presented in Algorithm \ref{alg:CBOAdaptive}.
Here, $R$ represents the number of moment refinement levels. For $R=0$ no refinement is performed, and a fixed moment approximation defined by $N_0$ moments is used.
\begin{algorithm}
\caption{Adaptive CBO method}\label{alg:CBOAdaptive}
\begin{algorithmic}
\Require $k$ reference data $U_{ref}(\lambda)$, $N_0<N_1<\dots<N_R$ moment approximations (sequence in $\mathbb{N}$) with tolerances $\epsilon_0>\epsilon_1>\dots>\epsilon_R$ (sequence in $\mathbb{R}_{>0}$), $K$ maximum number of iterations, $M$ number of particles, pseudo-time step $\Delta \tau$, $\delta$, $\sigma$, $\alpha$
\State $N \gets N_0$
\State $\epsilon \gets \epsilon_1$
\State $j \gets 1$
\For{$i=1,\dots,M$}
\State $X^i \gets $ random i.i.d initial state in $\mathcal{X}_{ad}$
\EndFor
\For{$k=1,\dots,K$}
\For{$i=1,\dots,M$}
\State Compute the solution of the RTE \eqref{eq:RTE} with $SP_N$ for each wavelength and source $q(X^i)$ 
\State Compute the weight $\omega_\alpha^f(X^i)$
\EndFor
\State Compute the consensus point $c_f^\alpha(X)$
\For{$i=1,\dots,M$}
\State $X^i\gets X^i-\Delta \tau \delta(X^i-c_f^{\alpha}(X))+\sqrt{\Delta \tau}\sigma \text{diag} (X^i(\tau)-c_f^{\alpha}(X))W_i$ 
\State $W_i$ i.i.d standard normal distributed in $\mathbb{R}^{d}$
\State $X^i\gets\text{argmin} \{\|X^i-y\|, y\in\mathcal{X}_{ad}\}$
\EndFor
\State $V \gets \frac{1}{M}\sum_{i=1}^M\| X^i-c_f^\alpha(X)\|_2 $
\If{$V<\epsilon_R$}
     \State \textbf{break}
\ElsIf{$V<\epsilon$ and $N=N_{j}$}
    \State $j \gets j+1$
    \State $\epsilon \gets \epsilon_j $
    \State $N \gets N_j$
\EndIf
\EndFor
\end{algorithmic}
\end{algorithm}
\subsection{Numerical Approximation}
As the solution of the $SP_N$ equations cannot be computed explicitly, a numerical approximation is required. 
Subsequently, the objective function is approximated by a midpoint quadrature resulting in the discrete  $\ell^2$ norm.
The numerical solution to \eqref{eq:SPN} is obtained using a staggered grid  developed in \cite{seibold2014starmap} accessible at \href{https://github.com/starmap-project}{GitHub}. The code uses pseudo-time stepping to solve the $SP_N$ approximation to the RTE \eqref{eq:RTE}. We provide zero initial data for the pseudo-time stepping algorithm and simulate until the $L^2$ norm of the (approximate) temporal derivative inside the phantom $\mathcal{P}$ is small, i.e. $\Vert \partial_t \phi_0\Vert_{L^2(\mathcal{P})}\leq \varepsilon.$
As previously stated, a perfectly matched layer approach is employed to address the vacuum boundary conditions. This enables the utilization of the preexisting boundary conditions in  \cite{seibold2014starmap} for the outer boundary of the layer, as they do not impact the solution.
The method is a second order approximation of \eqref{eq:SPN} in space (and pseudo-time) and we refer
to  \cite{seibold2014starmap} for additional implementational details.
\section{Numerical results}
In the numerical examples, we chose $\epsilon_R=10^{-2}$ and compared a full $SP_1$ ($R=0,\ N_0=1$), i.e. the DA, $SP_3$ ($R=0,\ N_0=3$) and an adaptive $SP_1$-$SP_5$ algorithm ($R=2,\ (N_0,N_1,N_2)=(1,3,5),\ (\epsilon_0,\epsilon_1,\epsilon_2)=(10^{0},10^{-1},10^{-2})$, abbreviated with $SP_A$ in the following) using Algorithm \ref{alg:CBOAdaptive}. 
Hence, the first two algorithms have a fixed number of moments and use one model each.
Both of these models have been used extensively in the literature for BLT \cite{gu2004three,chaudhari2005hyperspectral,jiang2007image,han2008bioluminescence,kuo2007three,lu2009spectrally,yang2015performance,liu2010evaluation,lu2009experimental,guo2013improved,kokhanovsky2013multi,klose2010vivo,liu2022multispectral,yu2021bioluminescence}, but higher-order models than $SP_3$ are usually not considered due to the trade-off between accuracy and computational cost.
The adaptive algorithm uses $SP_1$, then $SP_3$, and finally $SP_5$ to achieve a higher order while keeping the computational cost low.\\
In all numerical experiments, the stopping criteria for the pseudo-time stepping algorithm was $\varepsilon=$5e-03.
The CBO algorithm was being executed with $\delta=1, \sigma=1$ and a fixed number of $500$ particles and a pseudo-timestep of $0.1$ to solve the stochastic differential equation.
For $\alpha\to\infty$ the consensus point $c_f^\alpha(X)$ approximates $\text{argmin}_{i=1,\dots,M}f(X^i)$. Hence, the latter was used in the numerical algorithm.\\
To compare the obtained results, we computed the localization error (LE), i.e. the distance of the barycenters of each source, and the DICE score. For two sets $A$ and $B$ with $|\cdot|$ denoting the number of elements of a given set, the DICE score is defined as
\[DICE(A,B)=2\frac{|A\cap B|}{|A|+|B|}.\]
Note that in our case the area (2D) or volume (3D) of each source can be computed explicitly.
More accurate localization is correlated with low LEs and high DICE scores.

\subsection{Simulation verification}
\begin{figure}[ht!]
    \centering
        \begin{subfigure}[t]{0.4\linewidth}
    \includegraphics[width=\linewidth]{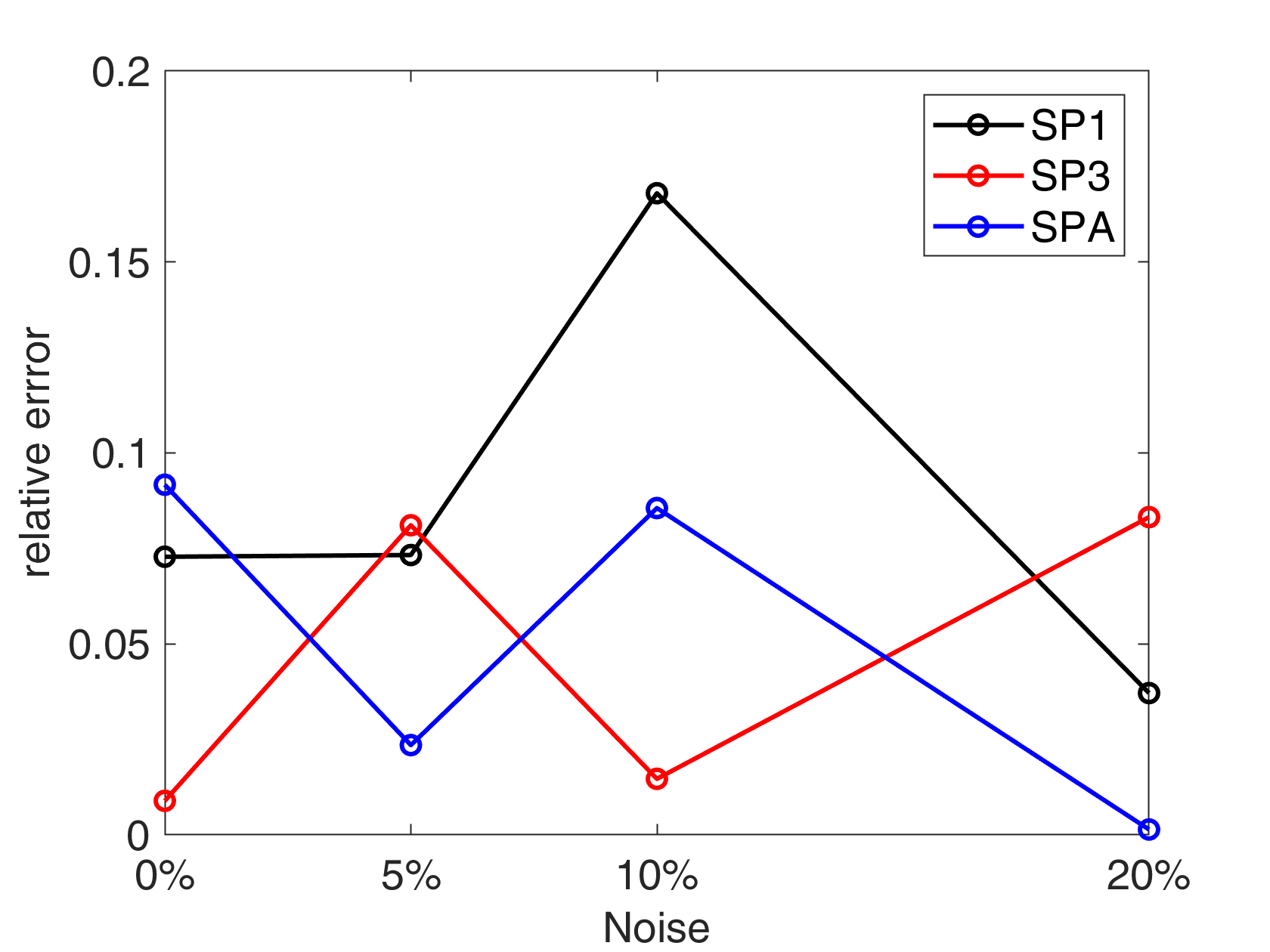}
    \caption{High scattering phantom (Table \ref{tab:absorptionscattering}).}
    \label{fig:noise1}
    \end{subfigure} 
        \begin{subfigure}[t]{0.4\linewidth}
    \includegraphics[width=\linewidth]{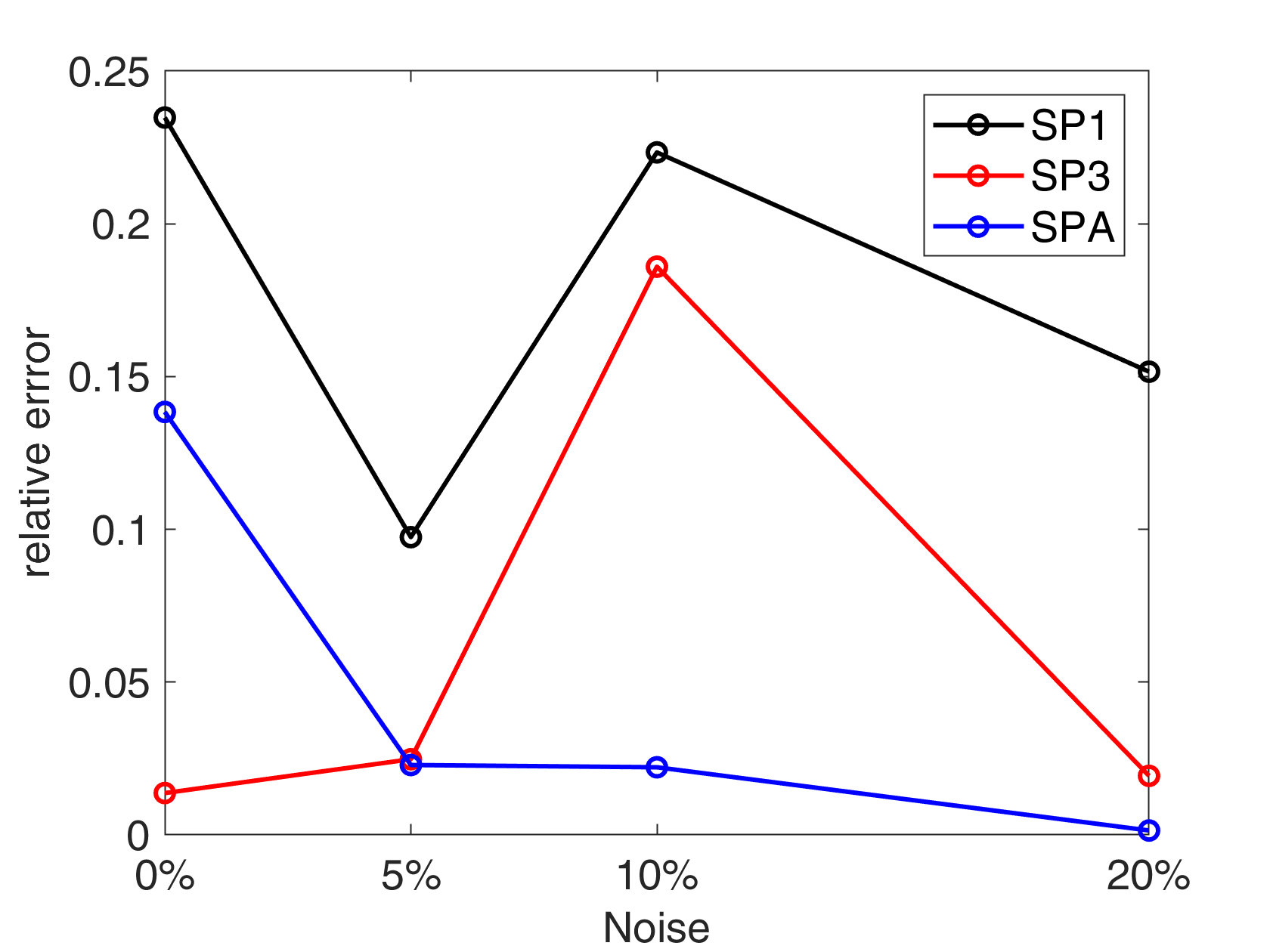}
    \caption{High absorption/moderately scattering phantom (Table \ref{tab:absorptionscattering2}).}
    \label{fig:noise2}
    \end{subfigure}
    \caption{Relative errors of the reconstructed source intensities in comparison to the real intensity for the source placed at $(0,0,0)$ and different levels of noise. SP1 represents the reconstruction with a full $SP_1$/DA algorithm and SP3 with $SP_3$, respectively. SPA denotes the adaptive CBO algorithm.}\label{fig:noisecomp}
\end{figure}

To verify Algorithm \ref{alg:CBOAdaptive} we set $\eta$ to zero and placed a solid spherical source with $0.5$ mm radius in a cube with an edge length of $7$ mm  at the coordinates $[-3.5,3.5]^3$. Three different positions of the source were investigated, $(0,0,-2)$, $(0,0,0)$, and $(0,0,2)$, at depths of $1.5$ mm, $3.5$ mm, and $5.5$ mm, respectively.
The reference solutions were a single view of the cube from the plane at $x_3=-3.5$ for each wavelength with $SP_{19}$.
For simplicity, the intensity of the source was constant over all wavelengths, such that each $q\in Q_{ad}$ is described by five unknowns: the intensity, the coordinates of the midpoint, and the radius of the sphere.
We used an equidistant grid with a grid size of $0.25$ mm in each direction.
The sets of absorption and scattering coefficients are given by Tables \ref{tab:absorptionscattering} and \ref{tab:absorptionscattering2} and corresponded to the values of the phantoms used in the experimental section, more details will be given there.
\begin{table}[ht!]
\centering
\caption{Absorption and scattering coefficients for the high scattering phantom.}
  \label{tab:absorptionscattering}
\begin{small}
\begin{tabular}{c|cccc}
\hline
$\lambda$ (nm) & 586 & 615 & 631 & 661 \\
\hline
$\mu_a$ (cm$^{-1}$)&1.1e-03  &2.678e-03 &2.916e-03 & 4.1e-03 \\
$\mu_s'$ (cm$^{-1}$) &{14.271}  &{13.523} &{13.129} &{12.425}  \\
\hline
\end{tabular}
\end{small}
   \vspace{-0.5cm}
\end{table}
\begin{table}[ht!]
\centering
\caption{Absorption and scattering coefficients for the high absorption/moderately scattering phantom.} \label{tab:absorptionscattering2}
\begin{small}
\begin{tabular}{c|cccc}
\hline
$\lambda$ (nm) & 586 & 615 & 631 & 661 \\
\hline
$\mu_a$ (cm$^{-1}$)&3.815  &3.569 &3.446 & 3.077 \\
$\mu_s'$ (cm$^{-1}$) &7.136  &6.762 &6.565 &6.213  \\
\hline
\end{tabular}
\end{small}
\end{table}
\begin{table}[ht!]
\centering
\caption{Results for the high scattering phantom (Table \ref{tab:absorptionscattering}).}
  \label{tab:Simuresults}
\begin{small}
\begin{tabular}{c|l|ccc}
\hline
Depth (mm) &Reconstruction & Coordinates & LE (mm) & DICE \\
\hline
&SP1 (DA)&(0.005,-0.010,-2.076)&0.077&0.884\\
1.5 &SP3&(-0.002,0.002,-1.930)&0.071&0.880\\
($x_3=-2$)&SPA&(-0.004,0.019,-1.942)&0.061&0.893\\
\hline
&SP1 (DA)&(0.006,-0.003,0.089)&0.089&0.864\\
3.5 &SP3&(0.005,0.007,0.094)&0.095&0.835\\
($x_3=0$)&SPA&(-0.002,-0.010,0.130)&0.131&0.762\\
\hline
&SP1 (DA)&(0.009,0.017,2.062)&0.065&0.715\\
5.5 &SP3&(-0.013,-0.015,1.711)&0.289&0.534\\
($x_3=2$)&SPA&(-0.013,0.005,1.902)&0.099&0.826\\
\hline
\hline
\end{tabular}
\end{small}
\end{table}
\begin{table}[ht!]
\centering
\caption{Results for the high absorption/moderately scattering phantom (Table \ref{tab:absorptionscattering2}).}
  \label{tab:Simuresults2}
\begin{small}
\begin{tabular}{c|l|ccc}
\hline
Depth (mm) &Reconstruction & Coordinates & LE (mm)& DICE \\
\hline
&SP1 (DA)&(-0.010,0.004,-1.867)&0.133&0.763\\
1.5&SP3&(-0.001,-0.015,-1.900)&0.102&0.847\\
($x_3=-2$)&SPA&(-0.025,-0.000,-2.120)&0.123&0.743\\
\hline
&SP1 (DA)&(-0.020,0.025,0.051)&0.060&0.910\\
3.5&SP3&(-0.008,-0.018,0.074)&0.076&0.885\\
($x_3=0$)&SPA&(0.008,-0.011,0.093)&0.094&0.772\\
\hline
&SP1 (DA)&(-0.015,0.042,1.630)&0.373&0.275\\
5.5&SP3&(0.073,0.0020,2.244)&0.256&0.597\\
($x_3=2$)&SPA&(0.014,-0.017,1.638)&0.362&0.351\\
\hline
\hline
\end{tabular}
\end{small}
\end{table}
\newline The localization of the $(x_1,x_2)$ coordinates was very accurate for all methods and both phantoms (Tables \ref{tab:Simuresults} and \ref{tab:Simuresults2}), since the reference data were provided in the $x_1$-$x_2$ plane.
For the high scattering example (Table \ref{tab:Simuresults}), all methods accurately reconstructed the sources with DICE scores above $0.72$ and LEs below $0.13$ mm, except for $SP_3$ at the deepest source.
The DA provided good results for all sources, probably due to the high number of scattering events.\\
In the high absorption simulation (Table  \ref{tab:Simuresults2}), the results for more shallow sources (depths $1.5$ and $3.5$ mm) are comparable to the high scattering phantom for all reconstruction methods: the DICE scores were above $0.74$ and the LEs below $0.14$ mm.
For all methods, the deepest source was reconstructed less accurately.
Nevertheless, at deep locations, the higher-order moment methods performed better than the DA, as reflected by the higher DICE scores and smaller LEs of $SP_A$ and $SP_3$ than for the DA.\\
\begin{figure}[ht!]
    \centering
    \begin{subfigure}[t]{0.4\linewidth}
     \centering
     \includegraphics[width=\linewidth]{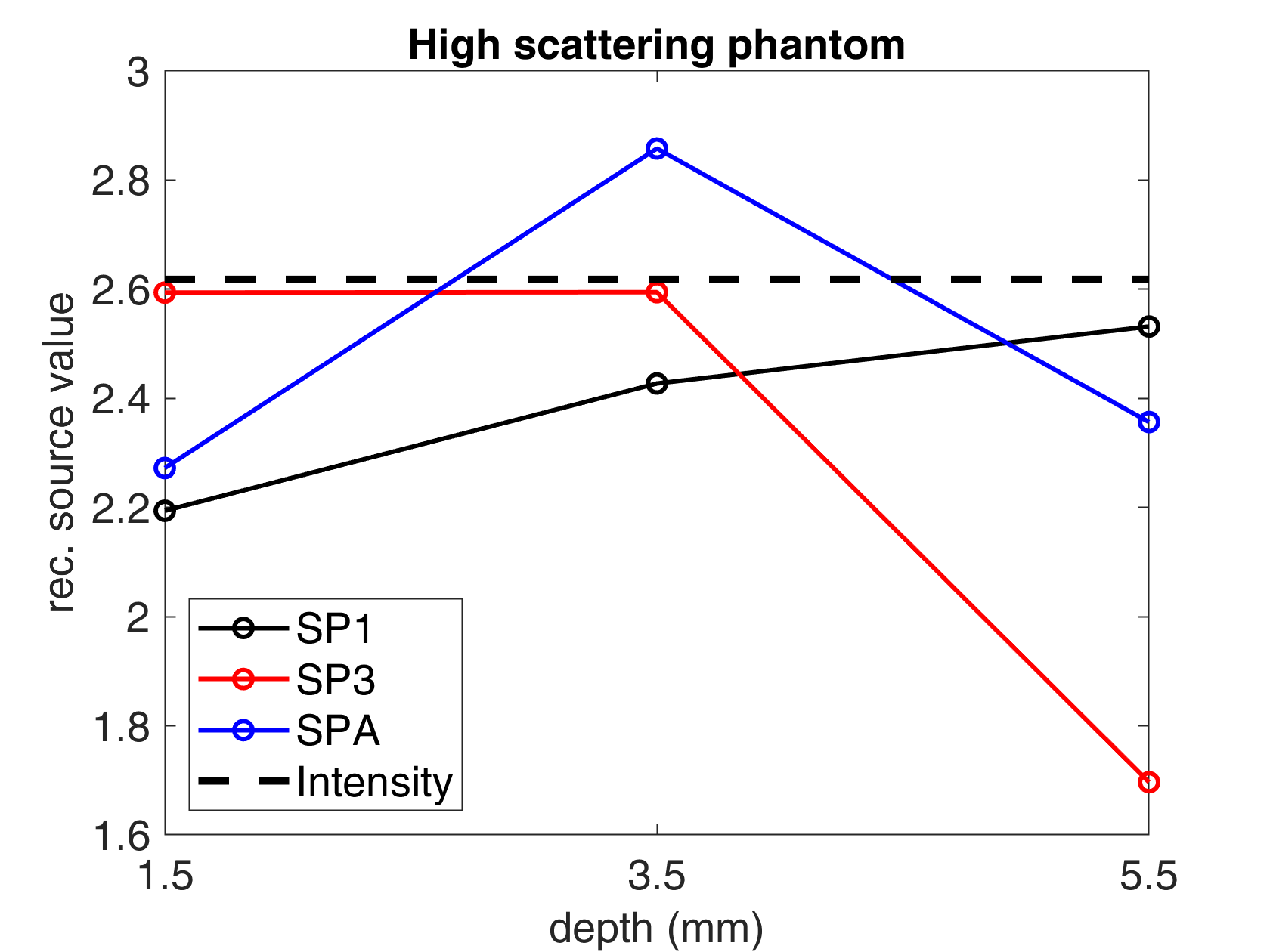}
     \caption{Reconstructed source values.}
     \label{fig:source4}
\end{subfigure}
    \begin{subfigure}[t]{0.4\linewidth}
     \centering
     \includegraphics[width=\linewidth]{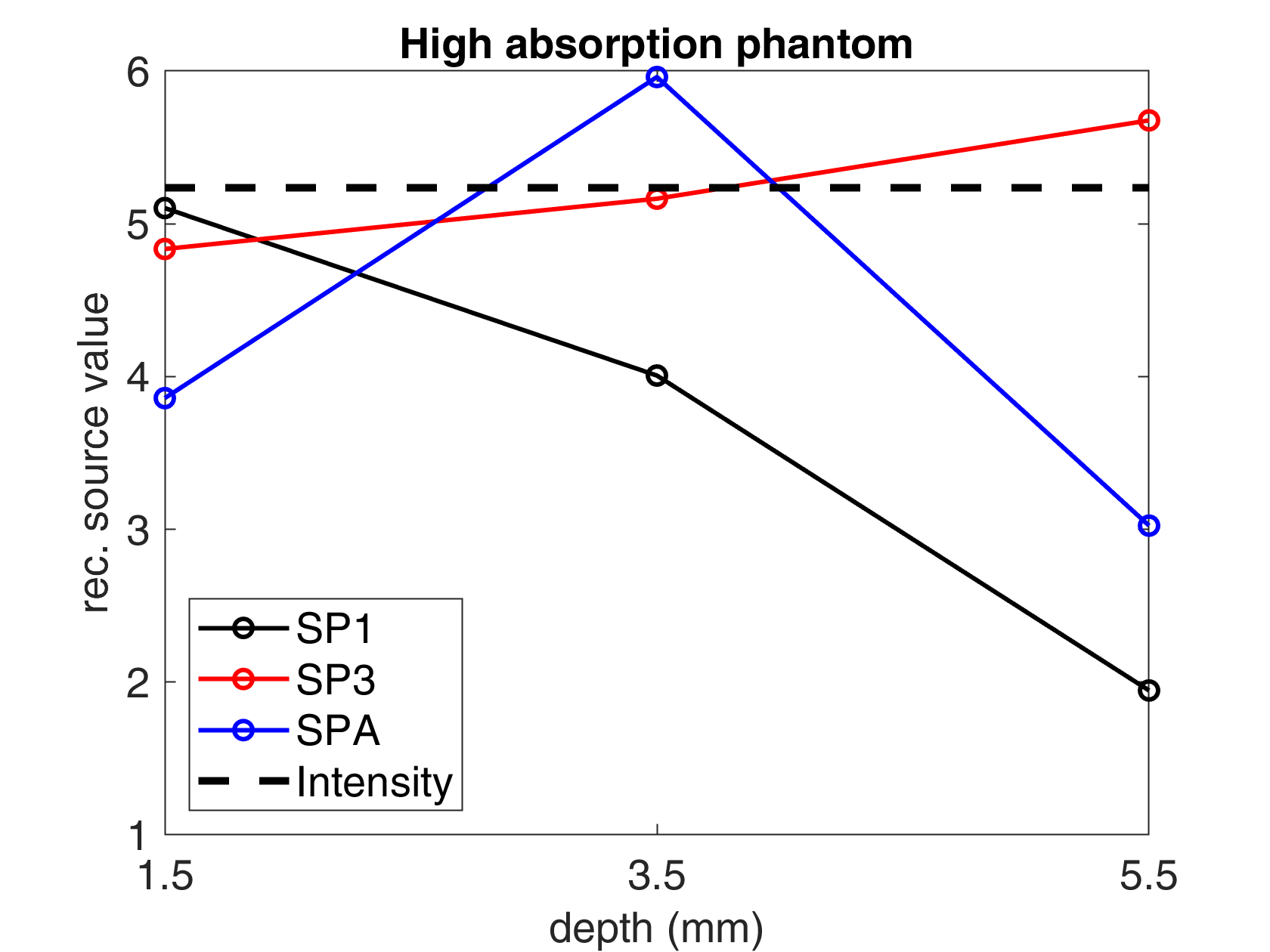}
     \caption{Reconstructed source values.}
     \label{fig:source1}
\end{subfigure} 
        \begin{subfigure}[t]{0.4\linewidth}
    \includegraphics[width=\linewidth]{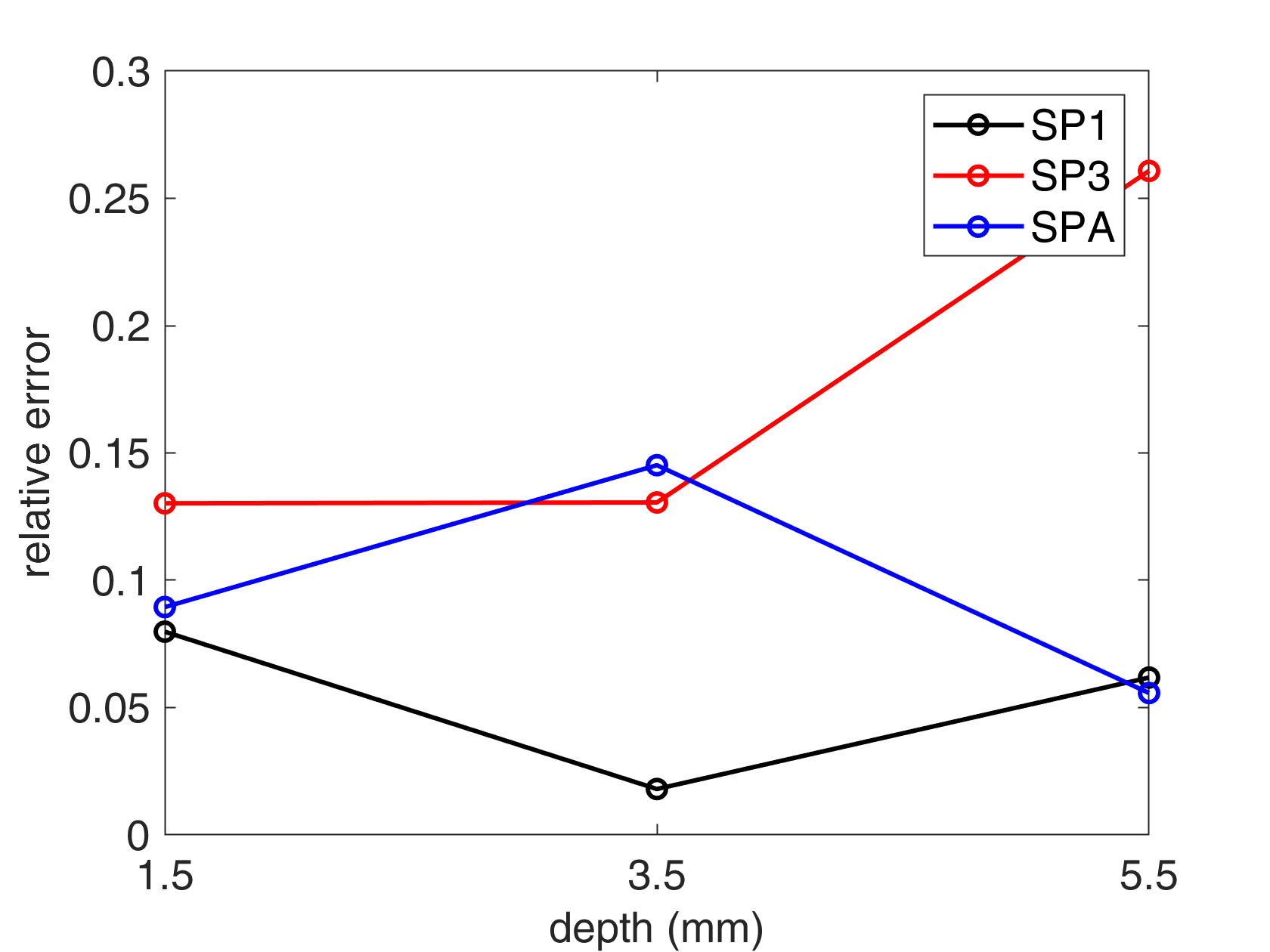}
    \caption{Relative error of the source values compared to the average source value of each reconstruction.}
    \label{fig:source5}
    \end{subfigure} 
        \begin{subfigure}[t]{0.4\linewidth}
    \includegraphics[width=\linewidth]{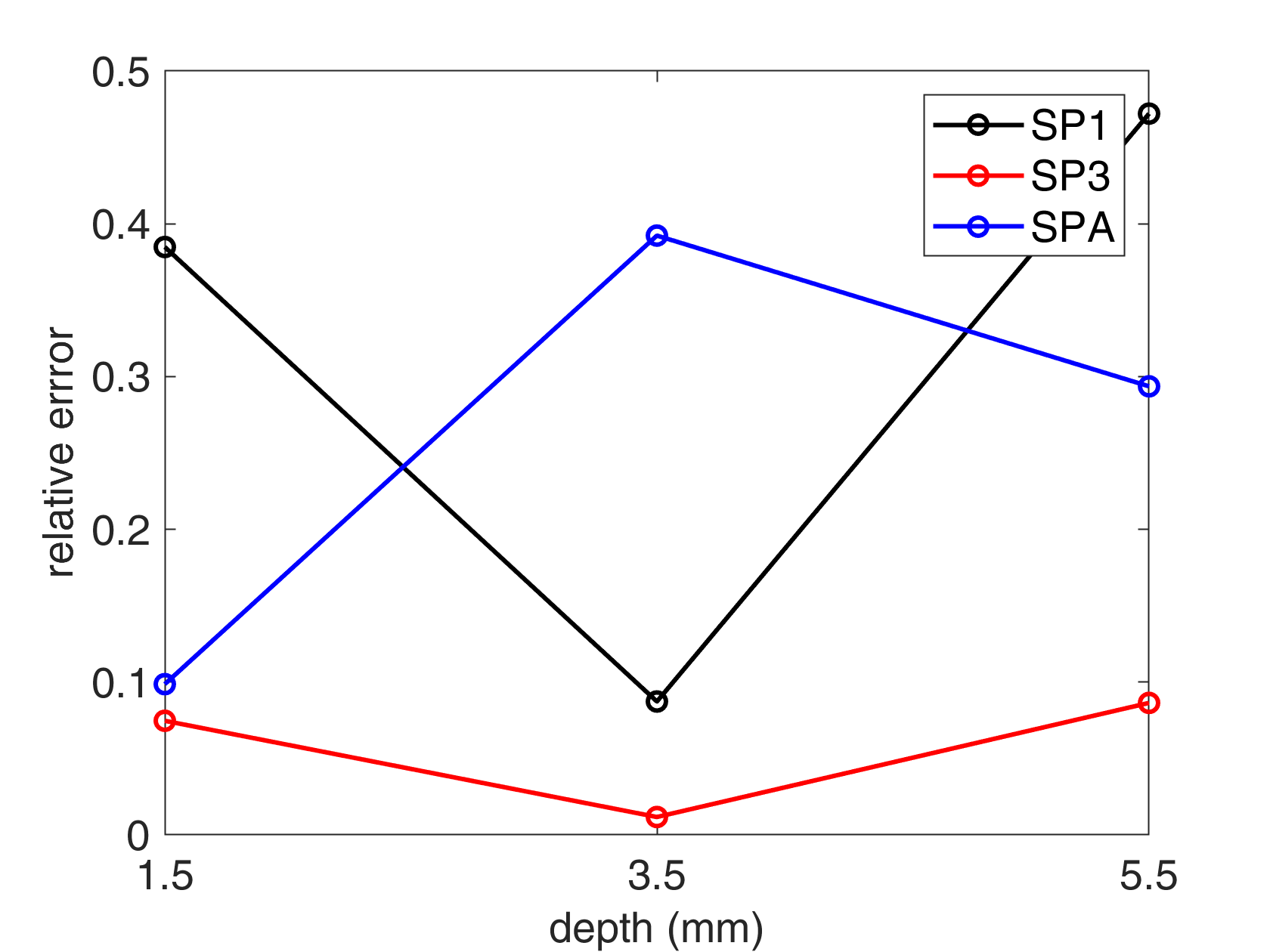}
    \caption{Relative error of the source values compared to the average source value of each reconstruction.}
    \label{fig:source2}
    \end{subfigure}
        \begin{subfigure}[t]{0.4\linewidth}
    \includegraphics[width=\linewidth]{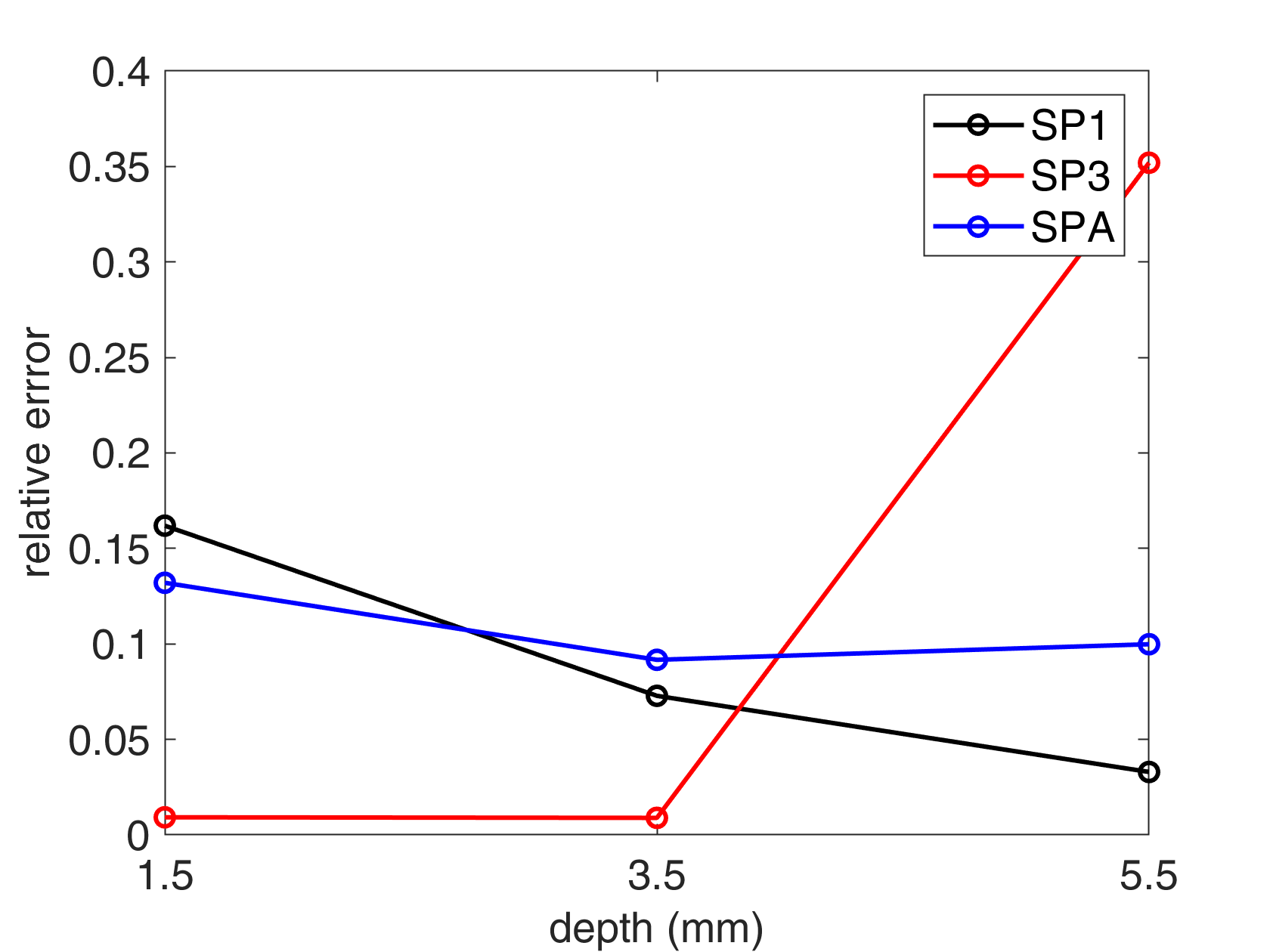}
    \caption{Relative error of the source values compared to the real source.}
    \label{fig:source6}
    \end{subfigure} 
        \begin{subfigure}[t]{0.4\linewidth}
    \includegraphics[width=\linewidth]{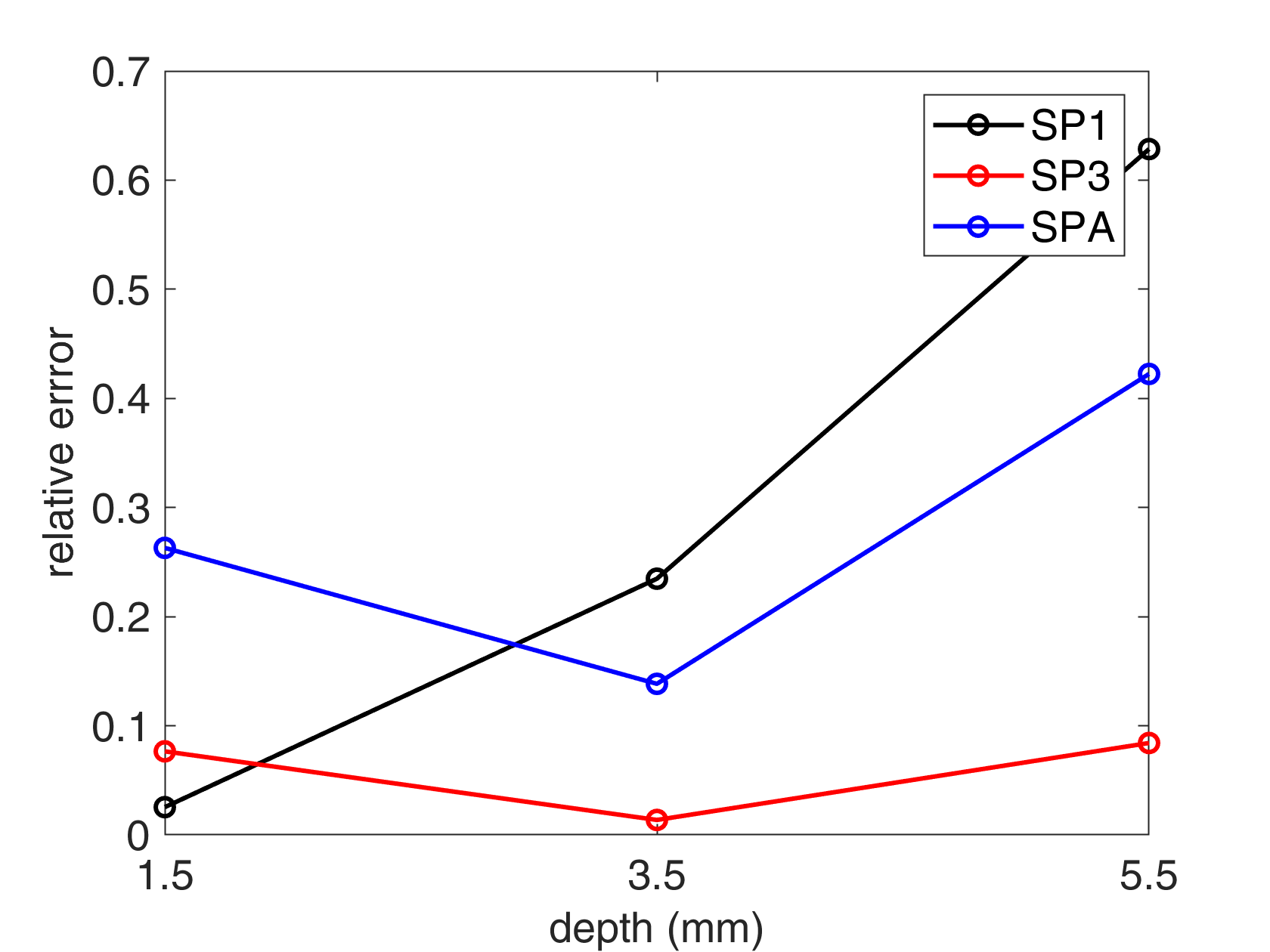}
    \caption{Relative error of the source values compared to the real source.}
    \label{fig:source3}
    \end{subfigure}
    \caption{Reconstruction of the source intensities for the source placed at different depths. SP1 represents the reconstruction with a full $SP_1$/DA algorithm and SP3 with $SP_3$, respectively. SPA denotes the adaptive CBO algorithm. Left column: The results for the high scattering phantom (Table \ref{tab:absorptionscattering}). Right column: The results for the high absorption/moderately scattering phantom (Table \ref{tab:absorptionscattering2}). \subref{fig:source4} and \subref{fig:source1}: Reconstructed intensities and real source intensity (dashed line), \subref{fig:source5} and \subref{fig:source2}: relative errors in comparison to the average source intensity of each reconstruction, \subref{fig:source6} and \subref{fig:source3}: relative errors in comparison to the actual source intensity.}\label{fig:sourcecomp}
\end{figure}
In addition to the localization, the reconstructed intensities of the different methods were compared, similar to \cite{lu2009spectrally}.
When the source is placed at different depths, the reconstructed intensities should be consistent.
The reconstructed source intensities are given by an integration over the entire reconstruction domain, see Figure \ref{fig:source4} and \subref{fig:source1}.
Further, we computed two relative errors: the relative error to the average source intensity for each reconstruction, see Figure \ref{fig:source5} and \subref{fig:source2}, and the error to the source used for the reference data, see Figure \ref{fig:source6} and \subref{fig:source3}.\\
In line with the previous results, the DA demonstrated to be an accurate and consistent reconstruction method for the high scattering phantom with relative errors 
mostly below 10\%, see \ref{fig:source5} and \subref{fig:source6}.
Note that the reconstructed values for the shallower sources of $SP_3$ were close to the real intensity, see \ref{fig:source4} and \subref{fig:source6}, but the deepest source was inaccurately reconstructed.
In the high absorption example, higher-order methods again performed better than the DA. 
Here, $SP_3$ was the most consistent method, see \ref{fig:source2}, and had small relative error terms (below 10\%, see Figure \ref{fig:source3}).
$SP_A$ performed better than the DA, too.\\
To test the performance of the proposed algorithms against noise, the source was placed in the middle of the square, i.e. at $(0,0,0)$, and reference data at $x_3=-3.5$ were disturbed by adding different levels of noise, i.e. $5\%$, $10\%$ and $20\%$.
The results for the high scattering example and for the high absorption example are given in Table \ref{tab:noise} and \ref{tab:noise2}, respectively.
In a highly scattering environment, LEs below $0.1$ mm were achieved for all reconstruction methods.
In addition, the DA still performed well in terms of DICE score, always above $0.8$, for all levels of noise.
For low and moderate noise levels the $SP_A$ achieved higher DICE scores than the DA.
For the high absorption example the higher moment methods outperformed the DA for a high level of noise, too.
The DICE scores were at least 0.14 higher and the LEs $0.12$ mm smaller than for the DA. 
In addition, the LEs of $SP_A$ for all noise levels were significantly lower than those of the DA.
\begin{table}[ht!]
\centering
\caption{Results for the high scattering phantom (Table \ref{tab:absorptionscattering}) with the source placed at $(0,0,0)$ and a different level of noise in the boundary data.}
  \label{tab:noise}
\begin{small}
\begin{tabular}{c|l|ccc}
\hline
Noise &Reconstruction & Coordinates & LE (mm) & DICE \\
\hline
&SP1 (DA)&(0.007,0.008,0.082)&0.082&0.874\\
$5\%$ &SP3&(-0.003,0.008,0.025)&0.027&0.951\\
&SPA&(-0.010,-0.030,0.022)&0.039&0.940\\
\hline
&SP1 (DA)&(0.008,-0.005,-0.066)&0.067&0.824\\
$10\%$&SP3&(0.017,-0.015,0.095)&0.097&0.812\\
&SPA&(0.046,-0.020,0.043)&0.065&0.898\\
\hline
&SP1 (DA)&(-0.003,0.005,-0.044)&0.045&0.802\\
$20\%$ &SP3&(0.007,0.001,0.081)&0.082&0.734\\
&SPA&(0.008,0.005,-0.035)&0.036&0.694\\
\hline
\hline
\end{tabular}
\end{small}
\end{table}
\begin{table}[ht!]
\centering
\caption{Results for the high absorption/moderately scattering phantom (Table \ref{tab:absorptionscattering2}) with the source placed at $(0,0,0)$ and a different level of noise in the boundary data.}
  \label{tab:noise2}
\begin{small}
\begin{tabular}{c|l|ccc}
\hline
Noise &Reconstruction & Coordinates & LE (mm) & DICE \\
\hline
&SP1 (DA)&(0.001,-0.007,0.149)&0.149&0.751\\
$5\%$ &SP3&(0.007,0.002.0.101)&0.101&0.845\\
&SPA&(0.002,-0.002,0.074)&0.074&0.717\\
\hline
&SP1 (DA)&(0.052,0.048,-0.005)&0.071&0.873\\
$10\%$&SP3&(-0.003,0.009,0.093)&0.093&0.755\\
&SPA&(0.012,-0.006,0.065)&0.067&0.881\\
\hline
&SP1 (DA)&(0.007,0.009,-0.176)&0.176&0.593\\
$20\%$ &SP3&(0.000,0.018,0.016)&0.024&0.735\\
&SPA&(-0.018,0.004,0.050)&0.052&0.923\\
\hline
\hline
\end{tabular}
\end{small}
\end{table}
\newline Additionally, the reconstructed source intensities were calculated for different noise levels.
The relative error can be seen in Figure \ref{fig:noisecomp}.
For the high absorption example, the reconstructed intensities were more accurate with higher--order methods, see \ref{fig:noise2}.
Even for the high scattering case, the relative errors were on average lower than those of the DA, see \ref{fig:noise1}.
\\
In summary, the examples demonstrate that the DA provides a very good localization of the source for different depths and noise levels in the case of a highly scattering phantom. However, high-order methods provide advantages when computing the actual source intensity in the case of noise in the reference data.
In the case of a highly absorbing phantom, the localization with higher-order moment methods against noise is even more superior to  DA.
Furthermore, source intensities (for different depths and against noise) calculated by higher--order moment methods are more accurate than by the DA for highly absorbing phantoms, which is relevant in vivo.

\subsection{Experimental reconstructions}
\subsubsection{Experimental set-up}
\begin{figure}[ht!]
    \centering
    \begin{subfigure}[t]{\linewidth}
     \centering
     \includegraphics[width=0.65\linewidth]{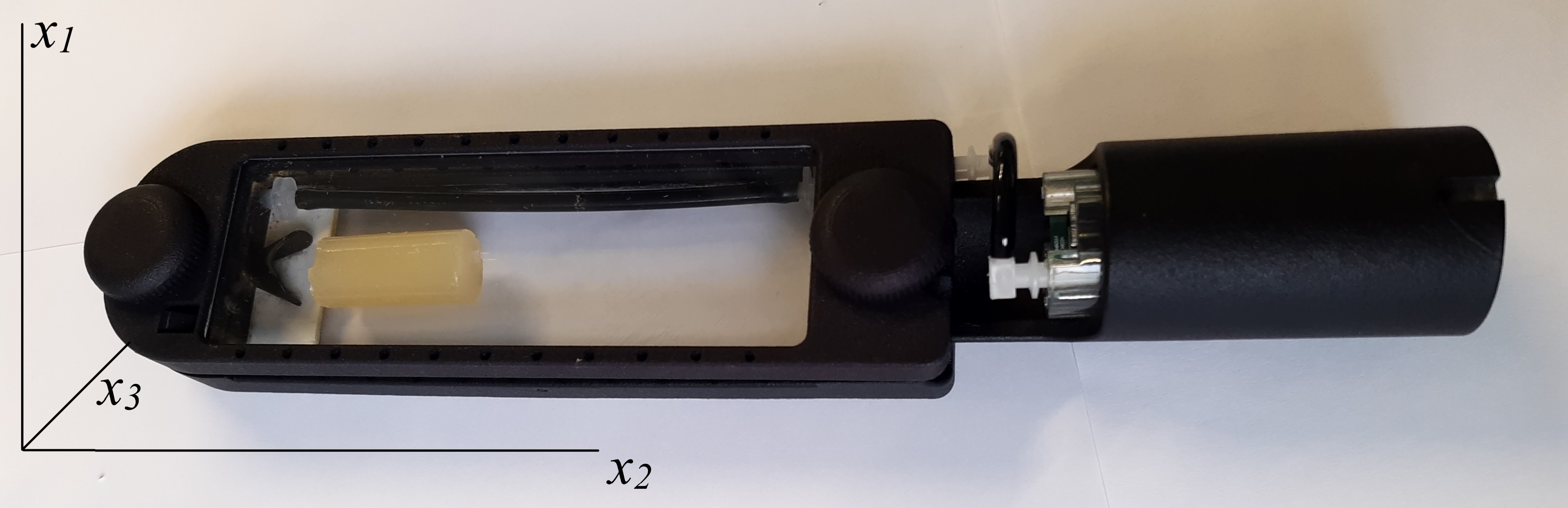}
     \caption{Gelatin phantom in the optical imaging holder. 
     }
     \label{fig:name}
\end{subfigure}
    \begin{subfigure}[t]{0.65\linewidth}
    \includegraphics[width=\linewidth]{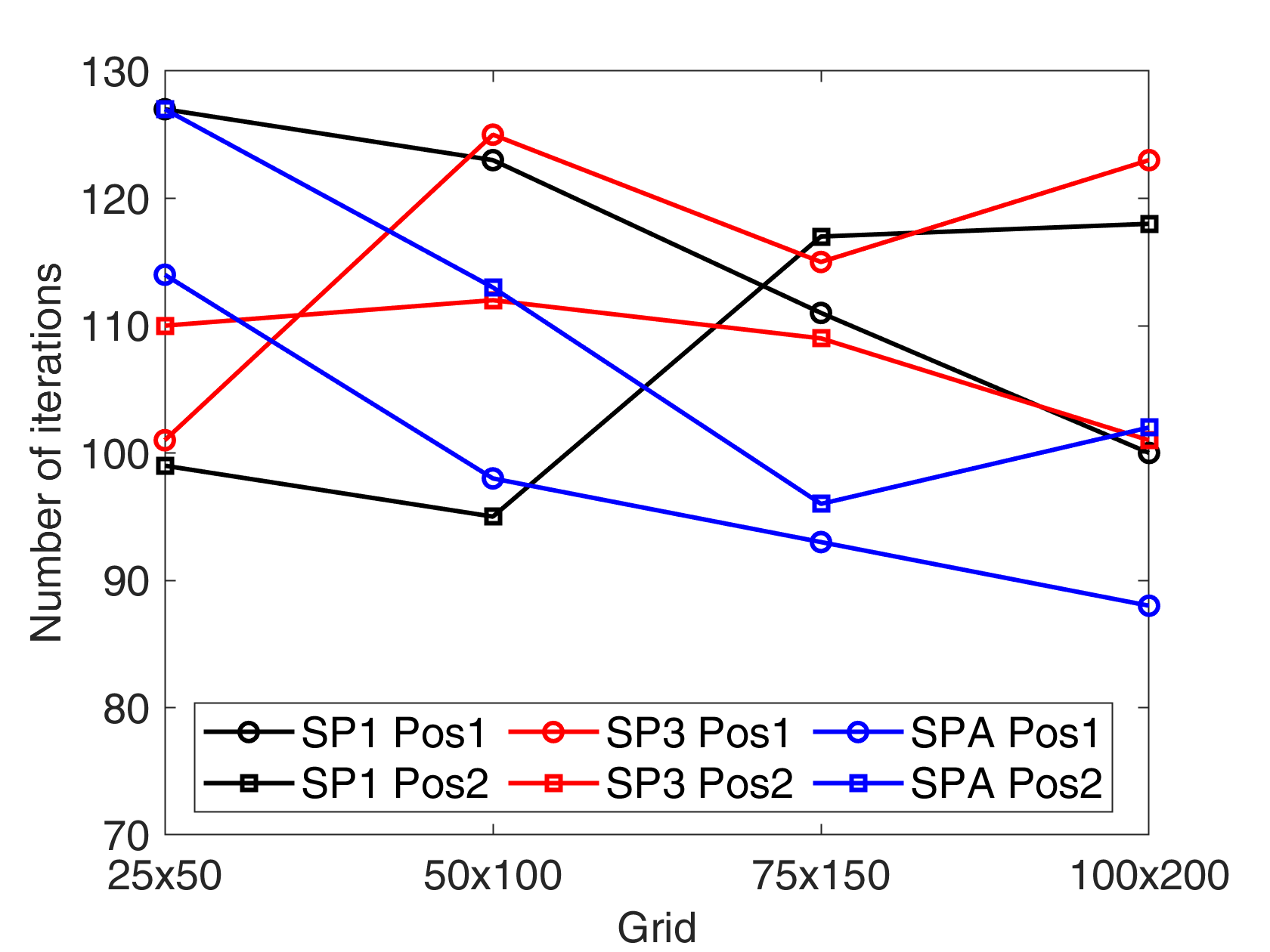}
    \caption{Number of iterations needed until the break criterion $\epsilon_R=10^{-2}$ is reached for different grids and moment approximations}
    \label{fig:gitterplot}
    \end{subfigure}    
    \caption{Experimental setup and computational effort: \subref{fig:name}: Cylindrical gelatin phantom placed in an optical holder. \subref{fig:gitterplot} The number of iterations needed until the break criterion $\epsilon_R=10^{-2}$ is reached for different grid sizes and moment approximations methods for the high scattering/low absorption phantom.}
\end{figure}

BLI was performed in a hybrid micro-CT optical imaging
system (MILabs B.V., Houten, the Netherlands), by capturing the emitted light with a charge-coupled device (CCD) camera at 586 nm, 615 nm, 631 nm, and 661 nm with an acquisition time of
approximately 2 minutes. After BLI, CT scans were performed in total
body normal scan mode, tube voltage of 55 kV, tube current of 0.17 mA,
an isotropic voxel size of 140 $\mu$m, and a scan time of 4 minutes. The reconstructed
BLI data had an isotropic voxel size of 280 $\mu$m.\\
The {low absorbing and highly scattering phantom} consisted of $20\%$ gelatin (reinst, Silber, 140 Bloom, Carl Roth, Mannheim, Germany) with $5\%$ Lipovenoes MCT $20\%$ infusion solution (Fresenius Kabi, Bad Homburg, Germany) in water, {while the second tissue-mimicking phantom contained only $2.5\%$ Lipovenös and 0.1 mg/mL of water-soluble Nigrosin (Thermo Fisher Scientific, Waltham, MA) resulting in lower scattering and higher absorption}. Phantoms were cast into a 15 mL tube (Sarstedt Inc, Nümbrecht, Germany). After solidification, the length has been adjusted to 3 cm, and a pipette tip containing 5 $\mu$L of the reactive solution of a red glow stick (Kontor3.11 GmbH, Chemnitz, Germany) was inserted into the phantom as described previously in \cite{schraven2023ct}. The phantom was scanned in two positions, with a rotation of about 90° (see Figure \ref{fig:name} for the phantom in the holder).
The absorption and scattering coefficients were calculated based on formulas derived in \cite{michels2008optical}.
{The low absorption phantom} was considered to consist of $5\%$ Lipovenoes MCT $20\%$ infusion solution (Fresenius Kabi, Bad Homburg, Germany) and $95\%$ water. 
As noted in \cite{michels2008optical}, Lipovenoes MCT $20\%$ is low absorbing at visible wavelengths.
For wavelengths larger than $500$ nm, the predominant absorber is water. The absorption of glycerin and egg lipids can be ignored due to the low concentration. The absorption of soybean oil {is low} for wavelengths larger than $500$ nm, hence, it can be ignored, too. The absorption coefficients for water can be found in \cite{pope1997absorption}.
The scattering coefficients were then approximated by \cite[Eq.  11 and Table 5]{michels2008optical} for Lipovenoes $20\%$ 
{ (see Table \ref{tab:absorptionscattering}).
For the high absorption/moderately scattering phantom the predominant scatterer is still given by Lipovenoes $20\%$ and neglecting the scattering of Nigrosin leaded to the scattering values of Table \ref{tab:absorptionscattering2}.
The predominant absorber is Nigrosin.
Here, the absorption coefficients were taken from \cite{AATBioAbsorption[Nigrosin]} and \cite{weber2011multispectral}.}
\subsubsection{Results}
\begin{figure}[ht!]
    \centering
    \includegraphics[width=0.5\linewidth]{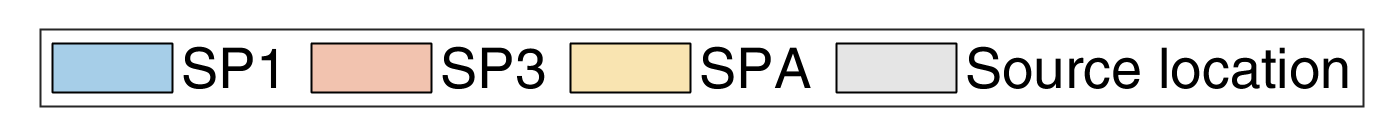}\\
    \begin{subfigure}[t]{0.24\linewidth}
     \centering
     \includegraphics[width=\linewidth]{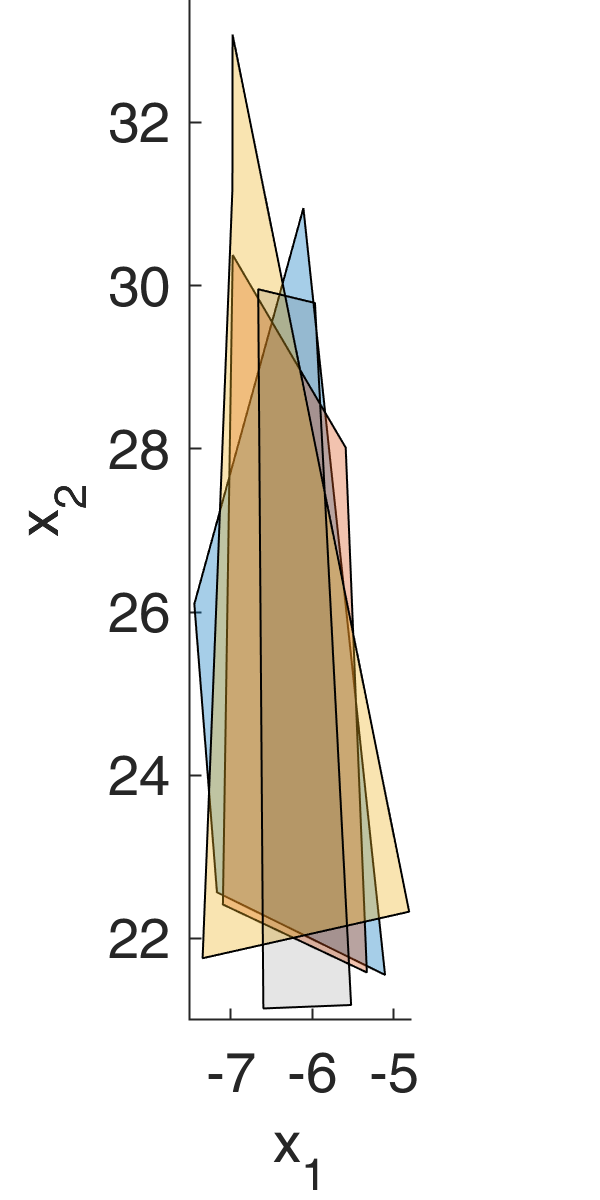}
     \caption{High scattering phantom, position 1}
    \end{subfigure}
    \begin{subfigure}[t]{0.24\linewidth}
     \centering
     \includegraphics[width=\linewidth]{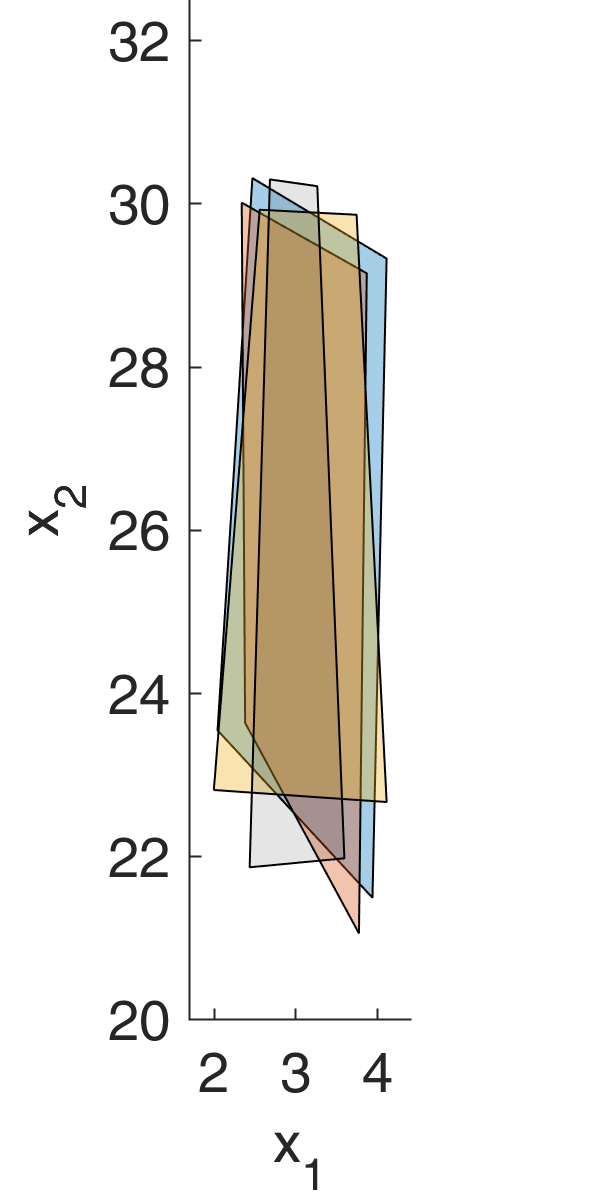}
     \caption{High scattering phantom, position 2}
    \end{subfigure}
    \begin{subfigure}[t]{0.24\linewidth}
     \centering
     \includegraphics[width=\linewidth]{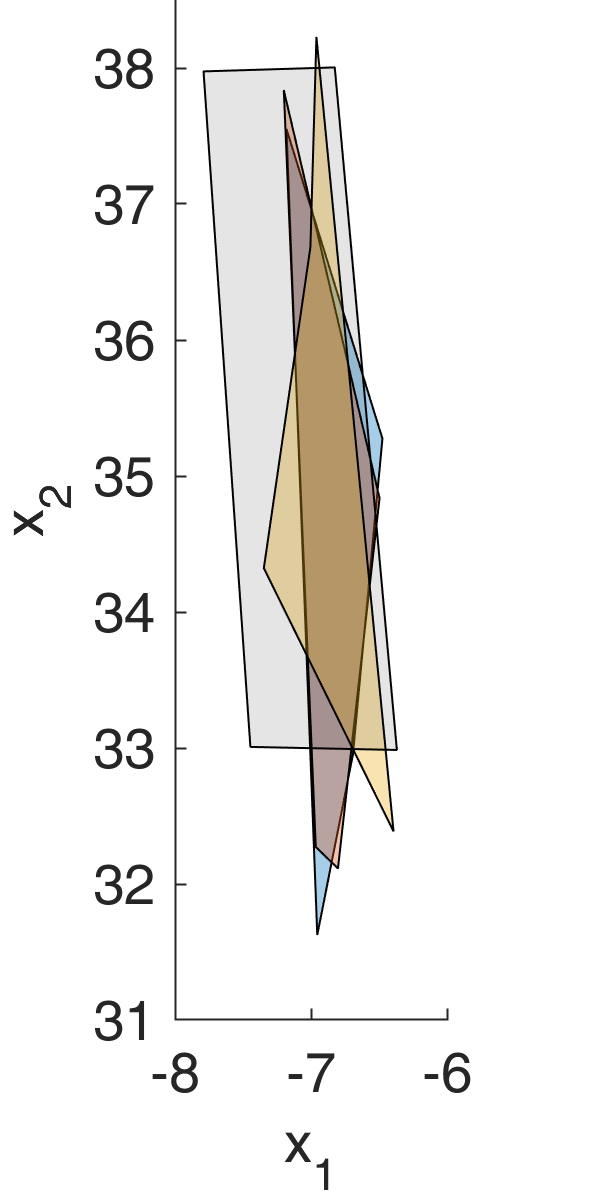}
     \caption{High absorption phantom, position 1}
    \end{subfigure}
    \begin{subfigure}[t]{0.24\linewidth}
     \centering
     \includegraphics[width=\linewidth]{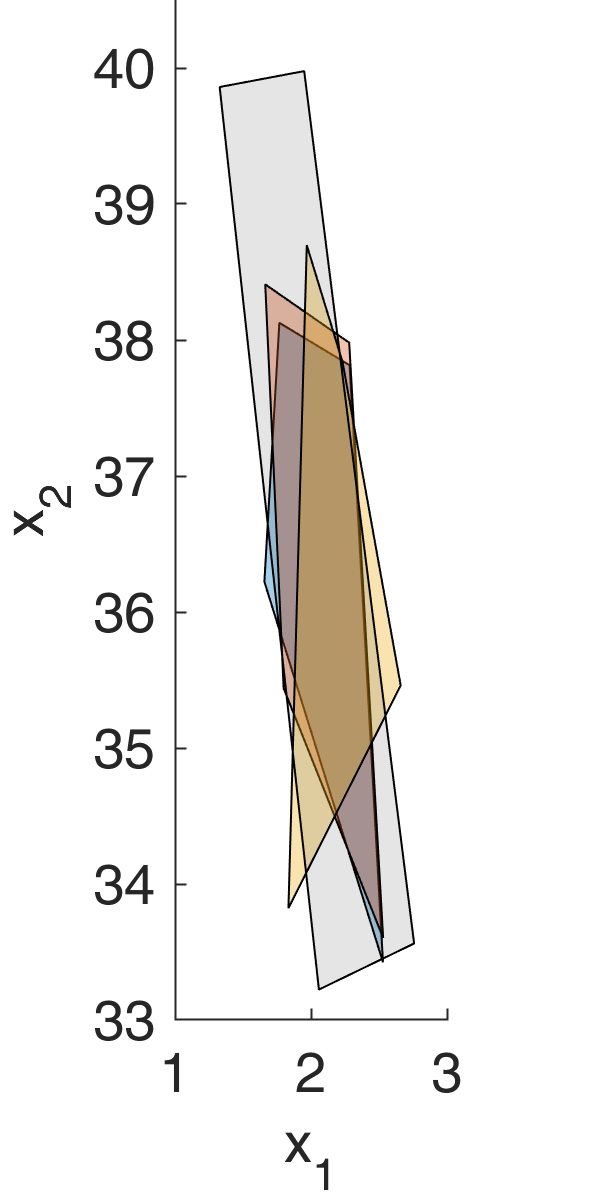}
     \caption{High absorption phantom, position 2}
    \end{subfigure}
    \caption{Reconstructed shapes of the sources computed by the different algorithms and the real source positions. Recall that the source shape was assumed to be quadrilateral. SP1 represents the reconstruction with a full $SP_1$/DA algorithm and SP3 with $SP_3$, respectively. SPA denotes the adaptive CBO algorithm. The real source locations were determined from CT data.}
    \label{fig:sourcerecons}
\end{figure}

The phantom had an approximate length of $3$ cm and a diameter of $1.5$ cm.
BLI data were recorded from the $x_1$-$x_2$ plane, where the boundary of the phantom $\partial \mathcal{P}$ is approximately a quadrilateral ($1.5$ cm by $3$ cm).
As previously described, a 3D computation is possible but rather expensive for smaller mesh sizes.
Therefore, we reduced the problem and assumed that $\Sigma_{a}$ and $\Sigma_s$ do not depend on the depth of the probe.
Then, the RTE was solved for  $\psi=\psi(x,\Omega)$ for $x \in \mathbb{R}^2$ reducing the computational overhead.
The inverse problem \eqref{eq:inverseproblem} can be solved in the $x_1$-$x_2$ plane, too.
The grid size was chosen to be $0.17$ mm in each direction.\footnote{This corresponds to a grid with 100 points in $x_1$ and 200 points in $x_2$ direction. Note that the computational domain included a small layer around the phantom to cope with the vacuum boundary conditions.}\\
The CBO directly solves the inverse problem, hence the number of iterations until reaching the break criterion does not depend on the mesh size.
This is visualized in Figure \ref{fig:gitterplot}, which displays the number of iterations for different grid sizes, for the high scattering/low absorption phantom, both positions, and the different algorithms. Note that the number of iterations for the high absorption/moderately scattering phantom displayed a similar behavior (data not shown).\\
Computed locations of the sources obtained with the DA, $SP_3$, and $SP_A$ were more accurate with higher moment methods than with DA (Table \ref{tab:meandist} and Figure \ref{fig:sourcerecons}). 
Note that Table \ref{tab:meandist} displays, in addition to the DICE score and LE, the objective function value, which can be seen as an approximation error of the solution. This value was always lower for the higher-order moment methods than the one of the DA.
For the highly absorbing phantom, this error decreased further for $SP_A$ and its fifth-order approximation.
Interestingly for the high scattering phantom, the reconstruction of the deeper source, i.e. position 2, was more accurate by $SP_A$ or $SP_3$ achieving a higher DICE score and lower LE than with the DA.
When the source was closer to the surface, DA and $SP_3$ gave the best results.
For the second phantom and position 1, the DA was still able to provide a similar LE and DICE score as $SP_3$, despite the high absorption.
Nevertheless, $SP_A$ outperformed the DA and $SP_3$.
For the second position $SP_3$ provided the most accurate results for the LE and DICE scores.
Figure \ref{fig:sourcerecons} shows that the reconstructed sources were larger than the actual sources for the highly scattering phantom and smaller for the highly absorbing phantom.
\begin{table}[ht!]
    \centering
        \caption{Localization error (LE) and DICE score and objective function value for different reconstructions}
    \label{tab:meandist}
    \begin{small}
    \begin{tabular}{l|l|ccc}
    \hline
    Dataset &Reconstruction & LE (mm) & DICE & $f_\eta(q(X^*))$\\
    \hline
    High scattering   &SP1 (DA)&0.243&0.668&1.924\\
    Position 1  &SP3&0.298&0.678&1.841\\
                &SPA&0.483&0.589&1.910\\
    \hline
    High scattering   &SP1 (DA)&0.460&0.626&1.267\\
    Position 2  &SP3&0.376&0.695&1.137\\
                &SPA&0.372&0.641&1.174\\
    \hline
    High absorption   &SP1 (DA)&0.795&0.485&0.584\\
    Position 1  &SP3&0.757&0.477&0.351\\
                &SPA&0.660&0.547&0.328\\
    \hline
    High absorption   &SP1 (DA)&0.374&0.616&0.454\\
    Position 2  &SP3&0.275&0.655&0.318\\
                &SPA&0.511&0.541&0.275\\
    \hline
    \hline
    \end{tabular}
    \end{small}
\end{table}
\newline In summary, all methods were able to reconstruct the sources for the data.
In all experiments a higher-order moment method gave the most accurate localization, even for the highly scattering phantom.
Concerning the localization of the source, there was no clear tendency between $SP_A$ and $SP_3$.
The approximation error was lower for higher-order moment methods, too.
In particular, the approximation error of $SP_A$ was the lowest for the highly absorbing phantom.

\section{Discussion}
The simulated and experimental setups demonstrate that CBO can be successfully applied in BLT. 
We were able to reconstruct unknown sources at different depths in the presence of noisy data using a hierarchy of $SP_N$ moment models.\\
In practice, the DA is often used to model the transmission of light \cite{gu2004three,chaudhari2005hyperspectral,jiang2007image,han2008bioluminescence,kuo2007three,liu2022multispectral}, so we used it to representatively model the state-of-the-art.
We found that the DA is an accurate model for highly scattering media.
Since the CBO method does not rely on gradient information, it was used for dynamically changing the number of moment equations, i.e. we increased the moments from $SP_1$ over $SP_3$ to $SP_5$ which achieves a high order while keeping computational cost low.
In particular, using higher-order methods in the last stage of the identification, showed an improved performance for deep sources or high absorption coefficients. 
This is significant for in vivo applications, since the high absorption/moderately scattering phantom of Table \ref{tab:absorptionscattering2} has similar optical properties (for the considered wavelengths) as, e.g. liver and spleen, see  \cite{mesradi2013experimental}. Furthermore,  in vivo,  light sources can be even at depths below 5.5 mm, for which improved depth registration by higher-order approximations together with CBO would be advantageous.
In addition, the total source intensities were more accurately recovered for higher--order methods in comparison to the state-of-the-art DA, particularly against noise. This becomes more relevant when the signal-to-noise ratio decreases, e.g. when small amounts of reporter gene-expressing cells have to be detected.
For the experimental data, there were also improvements using CBO in combination with higher-order moment approximations: The  error was lower and in all experiments,  the localization was more accurate, even for the highly scattering phantoms.
The adaptive $SP_A$ gave the lowest approximation errors for the highly absorbing phantom.\\
For computational reasons, we restricted our efforts to 2D without consideration of depth.
Computing a 3D solution on a similar mesh as the fine one in 2D ($100\times 200$ grid points) involves high computational cost, which additionally depends on the ratio of the scattering and absorption coefficients. 
For the simulated experiments in 3D and the DA ($SP_1$) the calculation of an optimal solution took approximately $52$ min (high absorption) and $118$ min (high scattering).
The computational cost increased by the factor  of $2.5$ (high absorption) and $3.5$ (high scattering) if $SP_3$ was used. Using a hierarchical CBO, the computational cost of $SP_A$ was comparable to $SP_3$, since the lower--order and cheaper approximations at the beginning of the algorithm, i.e. $SP_1$, balanced the higher--order approximations, i.e. $SP_5$ at the end.
Hence, we were able to achieve a higher--order approximation (i.e. $SP_5$) with comparable computational cost to a full $SP_3$ algorithm.
Due to the intrinsic parallelism of CBO, the computational 
time depends on the number of cores used. Numerical tests up to 24 cores indicated that the CBO algorithm is close to perfectly scalable. 
In contrast to the common applications of BLT, our approach is simplified to show the accuracy and possibilities of BLT reconstruction.  In vivo, the heterogeneity of the sample would be higher in terms of the optical properties of tissues and the potential for the light source to be located anywhere or at multiple points. The presented mathematical framework can however be directly applied to address these problems.
\\
In the future, we aim to further reduce the  computational effort of the inversion algorithm, by, for example, using suitable preconditioners similar to \cite{arridge2013preconditioning}, or by using a mini-batch method, similar to  \cite{carrillo2021consensus}. From an application point of view, the proposed method may also be extended to fluorescence tomography. 
This requires reconstructing the light path from the excitation laser to the fluorescence source (including absorption and scattering), autofluorescence, and additional information on absorption and location obtained from laser excitation - leading to a system of coupled approximations of the RTE equation. 
\\ 
In summary, CBO using hierarchical moment approximations
has been successfully applied in BLT thus providing  a promising new research direction  to improve preclinical optical imaging.

\paragraph{Funding}
Deutsche Forschungsgemeinschaft (DFG, German Research Foundation): Project-IDs 403224013 (SFB 1382) and 525853336, 525842915, 526006304 (all SPP2410); European Union’s Horizon Europe research and innovation programme under the Marie Sklodowska-Curie Doctoral Network Datahyking: Grant No. 101072546

\paragraph{Disclosures} The authors declare no conflicts of interest. 

\paragraph{Data Availability Statement}
Data underlying the results presented in this paper are not publicly available at this time but may be obtained from the authors upon reasonable request.

\bibliographystyle{siam}
\bibliography{sample}

\end{document}